\documentstyle[psfig]{mn}
\def\G{{\cal G}}
\def\ln{{\mathrm ln}}
\def\myr{$\,M_{\sun}\,$yr$^{-1}$}
\def\Myr{\,M_{\sun}\,{\mathrm yr}^{-1}}
\def\macc{$\dot M_{\mathrm{acc}}$}
\def\Macc{\dot M_{\mathrm{acc}}}
\def\lacc{$L_{\mathrm{acc}}$}
\def\Lacc{L_{\mathrm{acc}}}
\def\h2{${}^2$H}
\def\msun{$M_{\sun}$}
\def\Msun{M_{\sun}}
\def\rsun{$R_{\sun}$}
\def\Rsun{R_{\sun}}
\def\lsun{$L_{\sun}$}
\def\Lsun{L_{\sun}}
\def\teff{T_{\mathrm{eff}}}
\def\chem#1{${}^{#1}$}

\title{The Accretion of Brown Dwarfs and Planets by Giant Stars -- I. AGB Stars}
 
\author[Lionel~Siess and Mario~Livio]{Lionel~Siess$^{1,2}$ and
Mario~Livio$^1$\\ $^1$Space Telescope Science Institute, 3700 San Martin
drive, Baltimore, MD 21218 \\ $^1$Laboratoire d'Astrophysique de
l'Observatoire de Grenoble, Universit\'e Joseph Fourier, B.P.53X, F-38041,
Grenoble Cedex, France}
 
\begin{document}

\maketitle

\begin{abstract}
We study the response of the structure of an asymptotic giant branch (AGB)
star to the accretion of a brown dwarf or planet in its interior. In
particular, we examine the case in which the brown dwarf spirals-in, and
the accreted matter is deposited at the base of the convective envelope and
in the thin radiative shell surrounding the hydrogen burning shell. In our
spherically symmetric simulations, we explore the effects of different
accretion rates and we follow two scenarios in which the amounts of
injected mass are equal to $\sim 0.01$ and $\sim 0.1$\msun. The
calculations show that for high accretion rates ($\Macc = 10^{-4}$\myr),
the considerable release of accretion energy produces a substantial
expansion of the star and gives rise to hot bottom burning at the base of
the convective envelope. For somewhat lower accretion rates ($\Macc =
10^{-5}$\myr), the accretion luminosity represents only a small fraction of
the stellar luminosity, and as a result of the increase in mass (and
concomitantly of the gravitational force), the star contracts. Our
simulations also indicate that the triggering of thermal pulses is delayed
(accelerated) if mass is injected at a slower (faster) rate. We analyze the
effects of this accretion process on the surface chemical abundances and
show that chemical modifications are mainly the result of deposition of
fresh material rather than of active nucleosynthesis. Finally, we suggest
that the accretion of brown dwarfs and planets can induce the ejection of
shells around giant stars, increase their surface lithium abundance and
lead to significant spin-up. The combination of these features is
frequently observed among G and K giant stars.

\end{abstract}
 

\section{Introduction}

The discovery of several extra solar planets and brown dwarf candidates
around solar type stars (e.g. Mayor \& Queloz 1995, Butler \& Marcy 1996,
Cochran \& Hatzes 1997, Rebolo et al. 1995, 1996, Basri et al. 1996) has
advanced our understanding of the formation of planetary systems, but at
the same time has also raised new astrophysical problems.  Notably, some of
these planets are found to orbit very close to the central star (e.g. Mayor
et al. 1997) and their orbits may be tidally unstable. For example, Rasio
et al. (1996) showed that the orbital decay of 51 Peg's companion is
inevitable (although on a long timescale), and the authors conclude that
the planet will ultimately spiral-in into the star. Furthermore, the
engulfing of planets is certainly a possibility when the star evolves to
become a giant (e.g. Livio \& Soker 1984). The present paper deals
specifically with the reaction of a giant star ``swallowing'' a massive
planet or a brown dwarf. This work was also motivated by the fact that the
presence of a binary companion (even of a substellar mass) has often been
invoked to explain the axisymmetrical morphology of elliptical and bipolar
planetary nebulae (PNe) (e.g. Iben \& Livio 1993; Soker 1992, 1996a; Livio
1997). Notably, if the substellar secondary is not too close to the
primary, it will interact with the primary only on the upper AGB
(e.g. Soker 1995, 1996b).

In the course of its evolution, a main sequence star climbs along the red
giant branch and eventually (at a later stage) can also reach the
asymptotic giant branch (AGB). During the giant evolutionary phases, a
planet may accrete from the wind emitted by the giant or be entirely
engulfed by the giant's envelope. Depending on the initial orbital
separation, the planet or brown dwarf may be swallowed during the red giant
phase, along the AGB, or survive. Once inside the envelope, due to viscous
friction and also to tidal forces, a fraction of the orbital kinetic energy
of the planet is converted into thermal energy and, as a consequence, the
binary separation decreases. During this process, orbital angular momentum
is imparted to the envelope. The spiraling-in of a planet or a brown dwarf
inside the envelope of an AGB star has been studied in several papers
(e.g. Livio \& Soker 1984; Soker, Harpaz \& Livio 1984). Using simplifying
assumptions, these authors were able to follow the evolution of the brown
dwarf's mass subject to accretion and mass loss (evaporation). Their model
reveals the existence of a critical mass $M_{\mathrm{crit}}$ below which
the planet is evaporated or collides with the core. For the particular AGB
star model used, they found $M_{\mathrm{crit}} \sim 0.02$\msun, but these
results must be taken with caution since the physical processes involved in
these calculations were treated only approximately. On the observational
side, the orbital separations between close binary stars seem to indicate
that low-mass secondaries tend to have small orbital separations
(e.g. Tables 2 and 3 in Iben \& Livio 1993). A crude extrapolation of these
data to low-mass stars (Soker 1996b) suggests that secondaries with masses
of $\la 0.1$\msun\, will end up at orbital separations of $\la 1$\rsun. As
we will see in the next section, it is likely that at these depths inside
giants stars, the secondary will be tidally disrupted or
evaporated. Consequently, very low-mass secondaries will probably be
accreted onto the AGB star's core.

In the present paper we follow the evolution of an AGB star that accretes
matter from the destruction of a brown dwarf or a massive planet, using a
spherically symmetric stellar evolution code. In the next section, we
present the initial model for the AGB star and the general input
physics. In Section 3, we describe the numerical approach and in Section 4
we analyze the results of our computations. The effects of accretion on the
nucleosynthesis and surface abundances are presented in Section \ref{nucleo}
and a general discussion (Section \ref{discuss}) and summary follow
(Section \ref{summary}).

\section{Input physics and values of physical parameters}

To determine the exact location of the accretion process, we require a
knowledge of the structure of the AGB star and the colliding brown
dwarf. Where the brown dwarf will disrupt and dissipate can be estimated 
from the properties of the two objects involved in this scenario. In
this section we thus present our initial model for the AGB star and look
into the properties of the brown dwarf. With these data at hand, we will
try to characterize the accretion process as well as different quantities
associated with this phenomenon.

\subsection{The AGB star}

The initial AGB star model (taken from Forestini \& Charbonnel 1997) has an
age of $t_0= 4.74\,10^8$\,yr. Its initial radius, mass, effective
temperature and luminosity are equal to $R = 198\Rsun$, $M = 2.9\Msun$
(initially $M = 3.0\Msun$), $\teff = 3340$K and $L = 4395$\lsun,
respectively. The star has already undergone 2 thermal pulses and is in the
process of relaxing. The internal structure is composed of a very extended
convective envelope which accounts for more than 99\% of the radius.  Below
the massive convective zone ($M_{\mathrm{env}} \simeq 2.36$\msun), we find
a thin radiative shell surrounding the hydrogen burning shell (HBS). In the
HBS, H is mainly burned through the CNO cycle. Below the HBS is the
intershell where H is exhausted but where the temperature is too low to
allow for helium burning.  Below the intershell, helium is efficiently
burned through 3$\alpha$ reactions in the so-called helium burning shell
(HeBS). At the centre of the star, below the HeBS, the degenerate core is
composed mainly of ${}^{16}$O and ${}^{12}$C. The core is highly degenerate
and nearly isothermal. In Table \ref{tab1} we summarize the properties of
the AGB star. Note that the maximum temperature occurs off-centre due to
neutrino losses.
\begin{table}
\caption{Physical properties of the initial AGB  model}
\label{tab1}
\begin{tabular}{@{}lccccccc}
 & envelope  & HBS & HeBS  & core \\
$M_{\mathrm{top}}$ (\msun) &  2.900783 & 0.545262 & 0.541040  & 0.463803 \\
$M_{\mathrm{bot}}$  & 0.546783 & 0.544266 & 0.463803 & 0\\
$R_{\mathrm{top}}$ (\rsun) & 198 & $9.26\,10^{-2}$ & $1.92\,10^{-2}$ & $1.08\,10^{-2}$ \\
$R_{\mathrm{bot}}$  & 0.707 & $2.49\,10^{-2}$ & $1.08\,10^{-2}$ & 0 \\
$T_{\mathrm{top}}$ ($10^6$ K) & $3.34\, 10^{-3}$ & 16.2 & 78.8 & 192.2 \\
$T_{\mathrm{bot}}$  & 2.46 & 54.7 & 192.2  & 135.5\\
$\rho_{\mathrm{top}}$ (g\,cm$^{-3}$) & $5.785\, 10^{-9}$  & 0.292 & 158.4
&  $1.12\,10^5$ \\ 
$\rho_{\mathrm{bot}}$ & $5.14\,10^{-4}$ & 115.1 &  $1.12\,10^5$  & $2.44\,10^6$
\end{tabular}
\end{table} 

\subsection{The brown dwarf}

Brown dwarfs are compact, low mass objects ($0.007\Msun \la M_{\mathrm{bd}} \la
0.08$\msun) which never undergo efficient hydrogen burning. Their evolution
consists of a collapse phase, followed by deuterium burning (for the higher
masses) and degenerate cooling (e.g. Stevenson 1991, Burrows et al.
1993). Zapolsky and Salpeter (1969) determined a mass-radius relation for
this type of object, given approximately by the interpolation formula 
\begin{equation}
\frac{R_{\mathrm{bd}}}{\Rsun} \simeq 0.105 \Bigl(\frac{2X^{1/4}}{1+X^{1/2}}
\Bigr)^{4/3} \, 
\end{equation}
where 
\begin{equation}  
X = \frac{M_{\mathrm{bd}}}{0.0032\Msun}\ .
\end{equation}
Approximating the brown dwarf by a polytropic gaseous sphere with index
$n=1.5$ (characteristic of a non relativistic degenerate gas), one derives
for the central density, pressure and temperature the expressions
\begin{eqnarray}
\rho_c & \approx & 8.44 \Bigl(\frac{M_{\mathrm{bd}}}{\Msun}\Bigr)
\Bigl(\frac{R_{\mathrm{bd}}}{\Rsun}\Bigr)^{-3} {\mathrm\ g\,cm^{-3}} \\
P_c & \approx & 8.66\,10^{15} \Bigl(\frac{M_{\mathrm{bd}}}{\Msun}\Bigr)^2
\Bigl(\frac{R_{\mathrm{bd}}}{\Rsun}\Bigr)^{-4} {\mathrm\ dyne\,cm^{-2}} \\
T_c & \approx & 1.7\,10^7\,\mu_{\mathrm{bd}} \Bigl(\frac{M_{\mathrm{bd}}}{\Msun}\Bigr)
\Bigl(\frac{R_{\mathrm{bd}}}{\Rsun}\Bigr)^{-1}{\mathrm\ K}\ ,
\end{eqnarray}
where $\mu_{\mathrm{bd}}$ is the mean molecular weight, $M_{\mathrm{bd}}$ and
$R_{\mathrm{bd}}$ are 
the mass and radius of the brown dwarf, respectively. In Table \ref{tab2},
we present several quantities associated with a brown dwarf and a very
low-mass star of 0.015 and 0.1\msun, respectively.
\begin{table*}
\caption{Physical properties of accreting bodies}
\label{tab2}
\begin{tabular}{@{}lccccccc}
$M$ (\msun) & $R$ (\rsun) & $\bar \rho_{\mathrm{bd}}$ 
(g\,cm$^{-3}$) & $\rho_c$ (g\,cm$^{-3}$) &  $P_c\
{\mathrm (dyne\,cm^{-2}}$)  & $T_c$ (K) \\
0.015 & 0.095 & 24.6 & 147.6 & $2.4\,10^{16}$ & $1.9\,10^6$\\
0.10 & 0.067 & 468.2 & 2806.2 & $4.3\,10^{18}$ & $1.8\,10^7$ \\
\end{tabular}
\end{table*}

A straightforward comparison of the values in Tables \ref{tab1} and
\ref{tab2} indicates that the brown dwarf's mean density is of the order of
the density encountered in the HBS. It also reveals that brown dwarf
radii are of the order of the radius of the hydrogen burning shell.

\subsection{Times scales and orders of magnitude}

Given the properties of the AGB star and the brown dwarf, we can now
attempt to determine the locus of the brown dwarf dissipation and the
ensuing accretion process.

First, we estimate the Virial temperature of the brown dwarf, since this
represents the temperature above which the thermal kinetic energy of the
star can overwhelm the gravitational binding energy of the  brown
dwarf. This is given by
\begin{eqnarray} 
T_{\mathrm{V}} & \sim & \frac{\G\, M_{\mathrm{bd}} \,\mu_{\mathrm{bd}}
    \,m_H}{k\, R_{\mathrm{bd}}} \\ 
    & \simeq & 2.3\,10^6 \mu_{\mathrm{bd}} \Bigl(\frac{M_{\mathrm{
    bd}}}{0.01\Msun}\Bigr) 
    \Bigl(\frac{R_{\mathrm{bd}}}{0.1\Rsun}\Bigr)^{-1}\ {\mathrm K}\ .
\end{eqnarray}
Such temperatures, of the order of a few millions degrees, are encountered
at the base of the convective envelope (see Table \ref{tab1}). Therefore,
initially relatively massive brown dwarfs ($M \ga 0.01$\msun) will probably
be able to reach the bottom of the convective envelope without being
entirely evaporated. Note that we do not expect the trajectory of the
spiraling-in brown dwarf to be strongly affected by the convective motions
since its keplerian velocity is always supersonic (see e.g. Livio \& Soker
1984).

As the brown dwarf gets closer to the stellar core, tidal effects become
important and they can induce strong distortions to the brown dwarf's
structure, and eventually they can even lead to a total disruption of the
brown dwarf. The elongation stress at the brown dwarf centre can be
approximated by (e.g. Soker et al. 1987)
\begin{displaymath}
S = \int_0^{R_{\mathrm{bd}}}\rho_{\mathrm{bd}}\, \frac{d}{dR}
\Bigl(\frac{\G M}{R^2}\Bigr)\, l\, dl = 
\xi\, \frac{\G\, \bar \rho_{\mathrm{bd}} \, R_{\mathrm{bd}}^2\, M}{R^3}\
{\mathrm dyne\,cm^{-2}}\ , 
\end{displaymath}
\begin{equation}
\mbox{ } \\ 
\end{equation}
where $\bar \rho_{\mathrm{bd}}$ is the mean density of the brown dwarf, $M$
and $R$ the mass and radius of the star causing the disruption, and $\xi$
is a parameter of order unity. If we compare this expression with the
central pressure of the brown dwarf, then the condition of disruption, $S >
P_c$, reads
\begin{displaymath}
R <  \Big(\frac{\G \bar \rho_{\mathrm{bd}} R_{\mathrm{bd}}^2
 M}{P_c}\Bigl)^{1/3}\ \simeq\ 0.27\, \Bigl(\frac{M}{
 \Msun}\Bigr)^{1/3}\Bigl(\frac{\bar \rho_{\mathrm{bd}}}{100\,{\mathrm
 g\,cm}^{-3}}\Bigr)^{1/3} 
\end{displaymath}
\begin{equation}
\mbox { }\ \times \ 
 \Bigl(\frac{R_{\mathrm{bd}}}{0.1 \Rsun}\Bigr)^{2/3}
\Bigl(\frac{P_c}{10^{17}{\mathrm dyne\, m^{-2}}}\Bigr)^{-1/3}\ \Rsun\ .
\label{critere}
\end{equation}
From condition (\ref{critere}), it is clear that a disruption of the brown
dwarf occurs in regions close to the bottom of the convective
envelope. Thus, all the indications are that the brown dwarf will be
dissipated (and its material accreted) near the bottom of the convective
layer. 

In order to estimate the mean accretion rate onto the AGB star's core, we
evaluate the orbital decay rate $\tau_{\mathrm{decay}}$ in the accretion
region. The spiraling-in of the brown dwarf is due to the action of drag
forces that result from viscous friction and tidal effects. This
characteristic timescale, which is evaluated in the vicinity of the
dissipation zone, provides a rough estimates of the transit timescale of
the brown dwarf in this region and the rate at which matter can be
deposited there.  The accretion time scale is therefore given approximately
by (see e.g. Livio \& Soker 1984 for a detailed discussion)
\begin{eqnarray} 
\tau_{\mathrm{acc}} & = & \Bigl(\frac{\dot a}{a}\Bigr)^{-1} \simeq 162\,
 \Bigl(\frac{M_{\mathrm{bd}}}{0.01\Msun}\Bigr)^{-1} \Bigl(\frac{r}{0.7
 \Rsun}\Bigr)^{-3/2} \nonumber \\ & \times & \Bigl(\frac{M}{2
 \Msun}\Bigr)^{3/2} 
 \Bigl(\frac{\rho}{10^{-4}{\rm g\,cm^{-3}}}\Bigr)^{-1} {\rm yr}\ ,
\end{eqnarray}
where $\rho$ is the density in the envelope at radial distance $r$.
Consequently, the expected accretion rate \macc\, is of the order of $\Macc
\sim M_{\mathrm{bd}}/\tau_{\mathrm{acc}} \simeq 6\,10^{-5}$\myr for a
0.01\msun\, brown dwarf.

\subsection{The accretion process}
\label{sect}

The previous analysis indicates that a brown dwarf orbiting an AGB star can
spiral-in to very deep layers inside the star. On the basis of our
estimates of the Virial temperature and of tidal effects, we found that the
brown dwarf is likely to be dissipated at the base of the convective
envelope. The net outcome of the spiraling-in process is therefore the
accretion of a brown dwarf mass at rates in the range of $10^{-5}$ to
$10^{-4}$\myr.

\section{Numerical calculations}

\subsection{The stellar evolution code}

The stellar evolution code used in the present calculations has been
previously used for the computation of grids of standard pre-main sequence
(PMS) evolutionary tracks (Forestini 1994, Siess et al. 1997a), for the
detailed evolution and nucleosynthesis of intermediate mass AGB stars
(Forestini \& Charbonnel 1997), and also for the effects of disc accretion
onto PMS stars (Siess et al. 1997b, 1998). We refer the reader to these
papers for a detailed description of the stellar evolution code.

\subsection{The models}

We computed two sets of models for two different accretion rates, $\Macc =
10^{-4}$ and $10^{-5}$\myr\ (models A and B, respectively). The total
accreted mass in cases A and B was 0.01245\msun\, and 0.01238\msun,
respectively. The chemical composition of the accreted matter was assumed
to correspond to a solar metallicity ($Z = 0.02$) with relative abundances
from Anders and Grevesse (1989).

Previous works dealing with accretion in different evolutionary phases
(e.g. Stahler 1988, Siess et al. 1997b) have pointed out the important role
of deuterium in this process. Indeed, the very fast burning of \chem{2}H
through the \chem{2}H(p,$\gamma$)\chem{3}He reaction releases a huge amount
of energy which is able to modify the structure and evolution of the star
significantly. The presence or absence of deuterium in the accreted matter
is thus an important issue which depends essentially on the mass of the
brown dwarf. Brown dwarfs (or planets) more massive than $\sim 12$ Jupiter
masses ($\sim 0.01\Msun$, e.g. Burrows et al. 1995) have a high enough
central temperature to burn deuterium in their interiors. Consequently, we
do not expect deuterium to be accreted in the AGB star and its initial mass
fraction has been set to be $X_D = 10^{-15}$. We also assume that
\chem{7}Li is preserved in the brown dwarf and its mass fraction
(abundance) is equal to $X_{\mathrm{Li}} = 9.8968\,10^{-9}$
[log\,$\epsilon$(Li)=3.31].

\subsection{The deposition of accreted matter}

As we discussed in Section \ref{sect}, a sufficiently massive brown dwarf is
able to reach the bottom of the convective envelope. The deposition of some
accreted matter inside the convective envelope, due to partial evaporation,
will probably not in itself influence the structure since this region is
devoid of energy sources. The main effects that need to be studied,
therefore, are the consequences of mass deposition in the region close to
the bottom of the convective zone, where the planet/brown dwarf is expected
to disrupt.

To achieve this goal, we have assumed a simplified mass deposition profile
in which the mass $\Delta M_k = M_{k+1}-M_k$ of the shell $k$ is increased
at every time step $\Delta t$ by the amount $f\Delta M_k$, where $f = \Macc
\Delta t/(M_{\mathrm{top}}- M_{\mathrm{bot}})$. The mass is accreted in the
region delineated by $M_{\mathrm{bot}} < M_r \le M_{\mathrm{top}}$, where
$M_{\mathrm{bot}}$ corresponds to the mass coordinate at the top of the
HBS. However, if hot bottom burning is taking place (see below),
$M_{\mathrm{bot}}$ is equal to the bottom of the convective envelope (which
is inside the HBS). With this prescription, we allow the accreted matter to
go below the convective envelope, but, as we shall see later, a small
fraction of the accreted mass ($< 5\%$) is actually deposited in this
region. The boundary $M_{\mathrm{top}}$ is defined (arbitrarily) as the
mass coordinate corresponding to a radius equal to 1\rsun\
($M_{\mathrm{top}} \approx 0.55$\msun). This accounts for the fact that the
planet is actually dissipated in the central region of the star. We also
assume that the accreted matter is injected at the local envelope's
temperature and density. This can be justified by the fact that the brown
dwarf's kinetic energy is not too different from the specific thermal
energy of the AGB envelope at the same point (e.g. Harpaz \& Soker
1994). \\ In the presence of mass accretion, at each evolutionary
time-step, the total mass of the star is increased by the amount $\Delta M
= \Macc \Delta t$. In the shells where the mass is deposited (for
$M_{\mathrm{bot}} < M_r \le M_{\mathrm{top}}$), the time derivative of a
variable $A(M_r,t)$ is now calculated as
\begin{equation}
\frac{D A}{D t}\biggl|_{M_r}\, = \, \displaystyle \frac{\partial A}{\partial
t}\biggl|_{r} +\ (u.\nabla)A \ - \  \frac{d\, \ln M}{d t} \frac{\partial
A}{\partial\, \ln \bigl(\frac{M_r}{M} \bigr)} \Biggl|_{t} \ ,
\end{equation}  
where $u$ is the velocity of the shell and $\nabla = \frac{\partial
}{\partial r}$. The rate of gravitational energy release,
$\varepsilon_{\mathrm{grav}}$, is given by 
\begin{equation}
\varepsilon_{\mathrm{grav}} = - \frac{D e_{\mathrm{int}}}{D t} +
\frac{P}{\rho^2} \frac{D \rho}{D t}\ ,
\end{equation}
where $e_{\mathrm{int}}$ is the specific internal energy.

The chemical composition entering the computation at the new time step
is modified due to the arrival of fresh, not nuclearly processed material
from the accreted matter. The mass fraction $\tilde X(t+\Delta t)_{i,k}$ of
element $i$ at shell $k$ after mixing, is given by (Siess et al. 1997b)
\begin{equation}
\tilde X(t+\Delta t)_{i,k}\ =\ \frac{X(t)_{i,k} + f\,X^a_i}{1\,+\,f}\ ,
\label{xmchange}
\end{equation} 
where $X(t)_{i,k}$ is the stellar mass fraction of element $i$ before
mixing and $X^a_{i}$ is the mass fraction this element in the accreted
matter. Note that $X^a_{i}$ is constant and does not depend on the mass
coordinate.

Finally, in order to speed-up the convergence process, we turned off the
wind mass loss during the accretion phase and we used grey atmosphere
models.

\section{Evolution of the structure}
\label{struc}

The evolution of the AGB star under consideration can be divided into 2
major phases. The first one deals with the onset of accretion and the main
accretion phase while the second marks the end of the accretion process and
the relaxation of the star.

\subsection{The onset of accretion and the main accretion phase}

The deposition of material inside the star releases a significant amount of
gravitational energy. The associated accretion luminosity is given by
\begin{eqnarray}
\Lacc & = & \frac{\G M \Macc}{R} \approx 5650
\Bigl(\frac{M}{0.54\,\Msun}\Bigr) \Bigl(\frac{R}{0.3\,\Rsun}\Bigr)^{-1}
\nonumber \\ & \times &
\Bigl(\frac{\Macc}{10^{-4}\Myr}\Bigr)\,\Lsun\ .
\label{Elacc}
\end{eqnarray}
Note that an additional source of energy can arise from the possible
replenishment of the nuclearly active region with accreted material. The
modification of the chemical composition of these regions can increase the
nuclear energy production.

\begin{figure}
\psfig{file=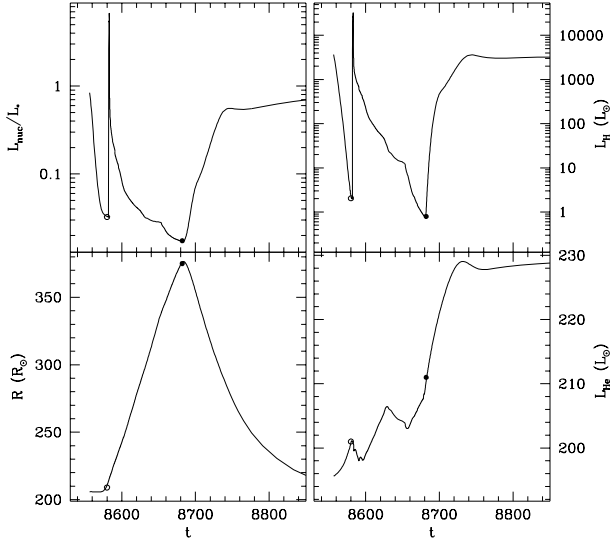,width=\columnwidth}
\caption[]{The evolution of the radius ($R$), nuclear luminosity
($L_{\mathrm{nuc}}$) and nuclear luminosities associated with H
($\mathrm{L_H}$) and He ($\mathrm{L_{He}}$) burning, in the case of high
accretion rate (case A). Open and filled dots mark the beginning of the hot
bottom burning and the end of the accretion process, respectively. In case
A, 0.01245\msun\, has been accreted in the star.}
\label{MY404_01}
\end{figure}

\begin{figure}
\psfig{file=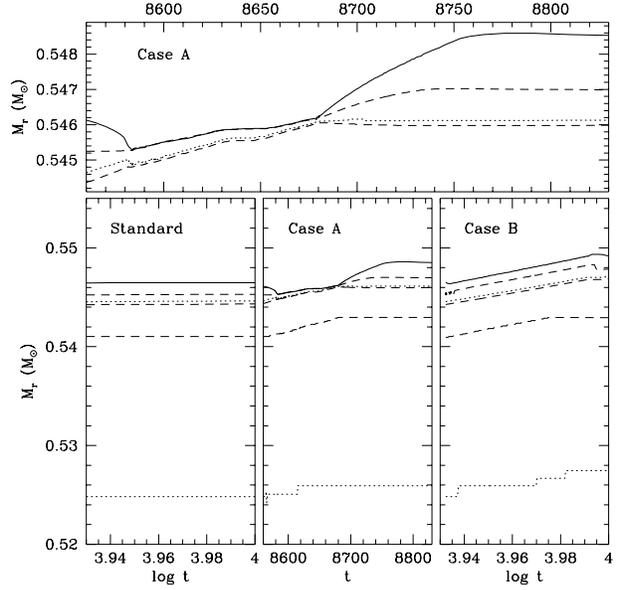,width=\columnwidth}
\caption[]{The evolution of the boundaries of the HeBS, HBS and convective
envelope. The dotted lines indicate the locus of maximum energy generation
while the dashed lines delineate the burning shells
($\varepsilon_{\mathrm{nuc}} > 5$ erg\,g${}^{-1}$s${}^{-1}$). Note that the
bottom of the HeBS has not been marked in the graphs, its location is at
about $M_r \simeq 0.47$\msun. The solid line indicates the lower edge of
the convective envelope. The evolution is presented for the standard case
(left panel) and for cases A and B (middle and right panels,
respectively). The important feature is the appearance of hot bottom
burning in case A. The top panel represents an enlarged view of the HBS in
case A.}
\label{MY404_02}
\end{figure}

\begin{figure}
\psfig{file=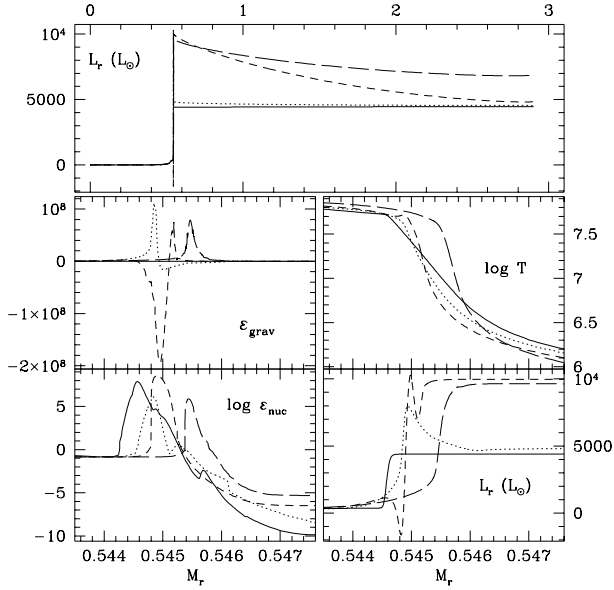,width=\columnwidth}
\caption[]{The evolution in the case of high accretion rates of the
profiles of gravitational (nuclear) energy production rates
$\varepsilon_{\mathrm{grav}}$ ($\varepsilon_{\mathrm{nuc}}$), temperature
($T$) and luminosity $L_r$ in the accretion region. The top panel shows the
profile of $L_r$ over the entire star. The solid curve refers to our
initial model, without accretion. The dotted, short- and long-dashed lines
correspond to models for which accretion is taking place for 15, 27 and 67
years, respectively.}
\label{MY404_03}
\end{figure}

For high accretion rates (case A), \lacc\ represents a large fraction of
the stellar luminosity and in the deep region where the energy is
deposited, the star expands. Locally, the density, pressure and temperature
decrease. The HBS, which advances in mass, rapidly reaches the ``cooler''
accretion region\footnote{For the following discussion, we identify the
accretion region as the region located above the HBS and up to a mass
coordinate $M_r \approx 0.55$\msun\ where the mass is deposited. Note also
that the top of the accretion region is located in the convective
envelope.} and then, the nuclear energy production due to H burning
decreases (see Fig. \ref{MY404_01}).
On the other hand, the luminosity of the HeBS increases, since the addition
of mass results in an increased gravity and compressional heating. In the
expanding accretion region, the opacity $\kappa$ increases since in this
regime it is a decreasing function of the temperature. The increase in both
$\kappa$ and $L$ (due to \lacc) contributes to a steepening of the
radiative gradient ($\nabla_{\mathrm{rad}} \propto \kappa L$) which now
becomes greater than the adiabatic gradient, thus convection sets in.
Consequently, the convective envelope penetrates rapidly towards deeper
regions (see Fig. \ref{MY404_02}) and reaches the HBS, leading to the
so-called hot bottom burning (HBB).
With the onset of HBB, the luminosity of the HBS ($\mathrm{L_H}$) rises
sharply, as fresh material coming from the convective region is engulfed in
this shell. The global contribution to the total nuclear energy production
of the \chem{13}C(p,$\gamma$)\chem{14}N reaction is increased during this
phase, and it accounts for 27\% (compared to 22\% in a standard
evolution). We also note that the \chem{14}N(p,$\gamma$)\chem{15}O and
\chem{15}N(p,$\alpha$)\chem{16}O reactions contribute less to the nuclear
energy production than in a standard scheme. These different features
indicate that the CNO cycle is reaching a new equilibrium. The ``flash''
induced by the penetration of the convective envelope into the HBS
increases $\mathrm{L_H}$ by 4 orders of magnitudes. The large release of
nuclear energy is used to expand the envelope and shortly after the
ignition of the HBB, $\mathrm{L_H}$ decreases rapidly
(Fig. \ref{MY404_01}). Thereafter, \lacc\ provides most of the
luminosity, the HBS almost extinguishes as it moves into the colder
accretion region, whereas $\mathrm{L_{He}}$ rises since the mass increases.
Figure \ref{MY404_03} depicts the profiles of several physical quantities
during the early accretion phase. The peaks in the luminosity profiles
indicate that the star is not in thermal equilibrium, but rather energy
accumulates in its interior. One also notices that this feature is smoothed
out as time advances, as a result of the star's expansion. In our initial
model (solid line), the gravitational energy production
$\varepsilon_{\mathrm{grav}}$ represents a small fraction of the stellar
energy production, 80\% of the luminosity is due to nuclear energy. When
the accretion process starts (dotted line), $\varepsilon_{\mathrm{grav}}$
increases rapidly below the accretion region as a result of mass
deposition.  In the accretion region where $\varepsilon_{\mathrm{grav}} <
0$, the expansion induces modifications to the temperature profile which is
now steeper. The nuclear energy production is consequently more localized
and concentrated to thinner regions. When HBB starts (short-dashed line),
both $\varepsilon_{\mathrm{grav}}$ and $\varepsilon_{\mathrm{nuc}}$ reach
extremum values.  The luminosity becomes negative, as most of the energy
production is used to expand the star. Finally, when the HBB is settled
(long-dashed line), the star heats up in the region of nuclear energy
production ($\varepsilon_{\mathrm{grav}}>0$), whereas expansion proceeds
above this zone. During the accretion phase the HBS narrows as the
temperature profile becomes steeper (Fig. \ref{MY404_02}). This also
accounts for the global decrease of $L_{\mathrm{nuc}}$ over this
period. Finally, since mass is being added, all the structural features
(location of the maximum energy production, burning shells, step in
temperature) move towards higher mass coordinates.

\begin{figure}
\psfig{file=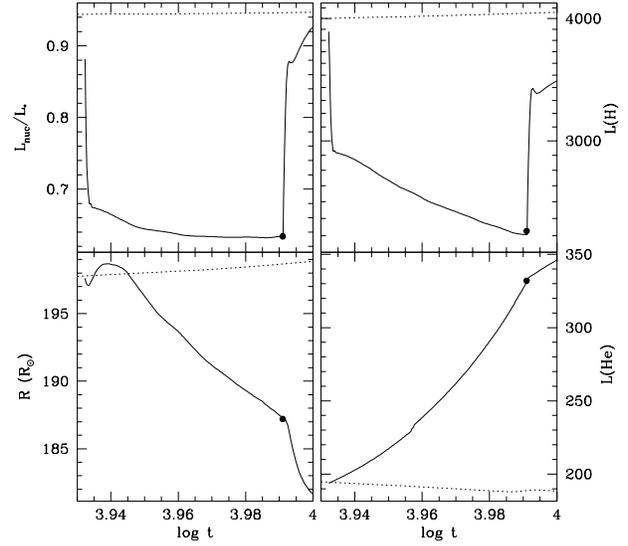,width=\columnwidth}
\caption[]{The evolution of the radius (R), nuclear luminosity ($L_{\mathrm{nuc}}$) and
nuclear energy production due to H ($\mathrm{L_H}$) and He ($\mathrm{L_{He}}$) burning. The solid
line refers to case B, the dotted line to the standard model. Filled
circles mark the end of the accretion phase.}
\label{MY404_04}
\end{figure}

In case B, the release of accretion energy is not sufficient to expand the
star and trigger convective energy transport down to the HBS. Consequently,
a thin radiative layer is maintained below the convective envelope, like in
the standard case (Fig. \ref{MY404_02}). This thin layer prevents the
energy production of the HBS from heating the envelope directly and, as
\lacc\ is small and the gravitational pull increases, the star
contracts. At the beginning of the accretion process, the release of
gravitational energy lowers the temperature in the accretion region (above
the HBS). As the core grows due to mass addition, the HBS moves outwards
and rapidly penetrates the less dense and cooler accretion region. As a
result, the nuclear luminosity due to hydrogen burning ($\mathrm{L_H}$)
undergoes a sudden drop (Fig. \ref{MY404_04}).
The star compensates for this decrease in energy production by a small
contraction. Thereafter, the star tries to find an equilibrium
configuration in which $L_{\mathrm{nuc}}$ and \lacc\ account for the total
luminosity. This phase of structural readjustment leads to a small
expansion. Thermal equilibrium is finally achieved after $\sim 130$\,yr,
which corresponds to the stellar Kelvin Helmholtz timescale
\begin{equation}
 \tau_{\mathrm{KH}} \simeq 150\, \Bigl(\frac{R}{200 \Rsun}\Bigr)^{-1}
\Bigl(\frac{L}{4500 \Lsun} \Bigr)^{-1}\Bigl(\frac{M}{3\Msun} 
\Bigr)^2 \ {\mathrm yr}\ .
\end{equation}
Subsequently, contraction resumes and the nuclear luminosity stays
at an almost constant value, somewhat lower than in the standard case
(without accretion), due to the presence of \lacc. However, the increase in
mass speeds-up the contraction rate in the central region and the
luminosity of the HeBS rises.\\  

\begin{figure}
\psfig{file=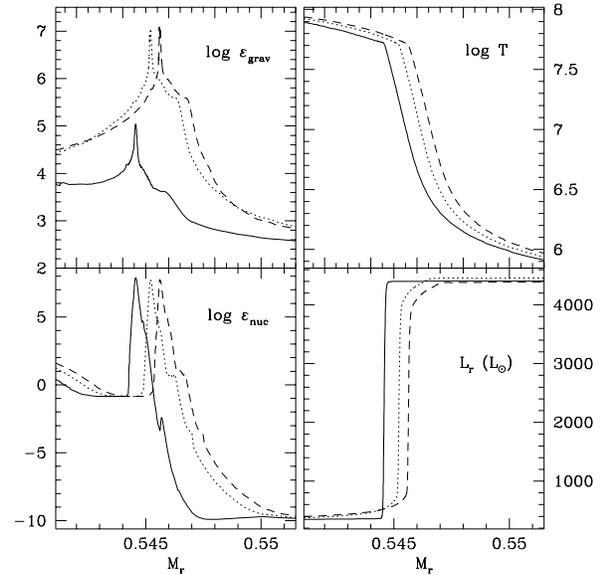,width=\columnwidth}
\caption[]{The evolution (in the case of low accretion rates) of the
profiles of luminosity ($L_r$), gravitational (nuclear) energy production
rates $\varepsilon_{\mathrm{grav}}$ ($\varepsilon_{\mathrm{nuc}}$) and
temperature ($T$) in the accretion region. The solid curves refer to our
initial model, without accretion. The dotted and short-dashed lines
correspond to models for which accretion is taking place for 327 and 535
years, respectively.}
\label{MY404_05}
\end{figure}

The main difference between case B and the standard evolution is the
persistence of stellar contraction, not only in the core but throughout the
entire structure during most of the accretion phase. For
low accretion rates, it is mainly the mechanical effects of mass deposition
and gravitational attraction that prevail. For a comparison with the high
accretion rate case, we present in Fig. \ref{MY404_05} 
the profiles of several physical variables during this period.
As can be seen, $\varepsilon_{\mathrm{grav}}$ remains largely positive in
the accretion region : contraction heats the central parts of the star. The
profiles of $T$, $\varepsilon_{\mathrm{nuc}}$ and $L$ are very similar
during the entire phase. We do not obtain a significant increase in the
steepness of the temperature gradient nor a peak in the luminosity
profile. The different quantities are simply shifted to higher mass
coordinates very smoothly. In this case, accretion does not lead to
significant structural changes, the energetic effects of accretion (\lacc)
being too weak.

\subsection{The relaxation of the star and the subsequent evolution}
\label{Sect1}

\begin{figure}
\psfig{file=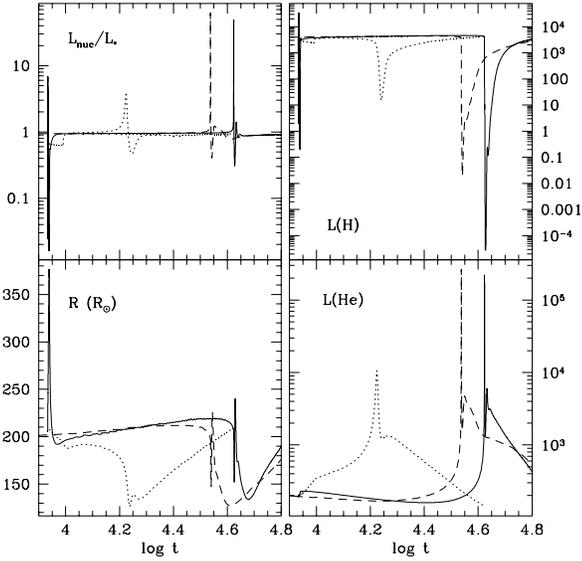,width=\columnwidth}
\caption[]{Relaxation of the star. The variables are the same as in 
Fig. \ref{MY404_04}. The solid, dotted and dashed lines refer to cases A,
B and to the standard evolution models, respectively.}
\label{MY404_06}
\end{figure}

For high accretion rates (Fig. \ref{MY404_01}), the sudden suppression of
accretion energy deposition initiates a rapid contraction of the star, as
the envelope can no longer be supported. The release of gravitational
energy increases the temperature in the regions of energy production and
induces an enhancement of the nuclear luminosities. The star globally heats
up, the opacity drops, and the convective envelope recedes towards the
surface (Fig. \ref{MY404_02}). The cessation of the accretion process thus
marks the end of HBB and this leads to the formation a thin radiative layer
between the convective zone and the HBS. The contraction proceeds as long
as the energy losses from the surface are not equilibrated by the gain from
gravitational and nuclear energy sources. The return to equilibrium is
finally achieved after $\sim 675$\,yr, when the stellar radius and
luminosity have decreased by a factor of $\sim 2$ (Fig. \ref{MY404_06}).
At this point, the nuclear luminosity provides more than 90\% of the
stellar luminosity and it is sufficient to support the envelope. As most of
the nuclear energy is produced by the HBS, the energetic influence of the
HeBS is shielded and its luminosity ($\mathrm{L_{He}}$) decreases (as in
the standard case). Finally, as the core mass $M_{\mathrm{core}}$ grows,
the temperature inside the HeBS increases and the nuclear reactions (which
are very temperature sensitive) runaway and trigger a thermal instability.
Note that in case A, the thermal pulse is delayed with respect to the
standard model. This is due to the large expansion that occurs during the
accretion phase, which in turn slows down the ``natural'' temperature
increase of the HeBS substantially.

In case B, when the accretion process stops, the star undergoes a short
phase of rapid contraction with the disappearance of \lacc.  The
temperature in the HBS increases and the nuclear luminosity rises
(Fig. \ref{MY404_06}). For a short period of time the star
expands. Finally, when an equilibrium configuration is reached, which
accounts for the missing \lacc, the star resumes contraction and
$\mathrm{L_{He}}$ keeps rising. The tendency of increasing
$\mathrm{L_{He}}$ is due to the fact that for low accretion rates the
impact of accretion is more a consequence of the mechanical effects
resulting from mass deposition (compressional heating), rather than
energetic effects due to \lacc\ or HBB. Note in addition that the accretion
rate used in this study is larger that the ``normal'' core growth rate
($\sim 10^{-7}$\myr, for a 3\msun\, star). Therefore, accretion forces the
star to evolve more rapidly than in a standard evolution and the thermal
pulse occurs earlier. However, the thermal instability is weaker in the
star accreting the brown dwarf than in the standard model: The maximum of
$\mathrm{L_{He}}$ is more than an order of magnitude lower than in the
conventional computation. The reason for this difference is the fact that
due to the mass deposition, the HeBS moves towards regions of a lower
density faster than in the standard scheme, and the nuclear energy
generation is thus reduced.

\subsection{Accretion of more massive bodies}

\begin{figure}
\psfig{file=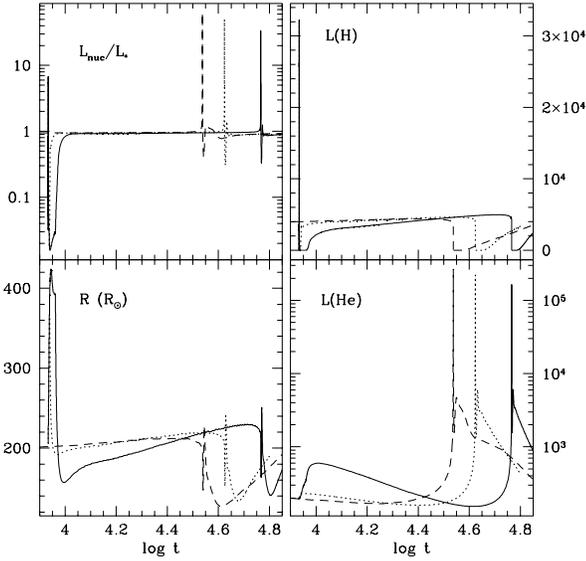,width=\columnwidth}
\caption[]{The influence of mass deposition for high accretion rates
($10^{-4}$\myr). The physical quantities are the same as in
Fig. \ref{MY404_04}. The solid curves correspond to a long-lasting
accretion phase ($M_{\mathrm{acc}} = 0.0594$\msun), the dotted line to case
A ($M_{\mathrm{acc}} = 0.01245$\msun) and the dashed line is the standard
model.}
\label{MY404_07}
\end{figure}

If a high accretion rate ($\Macc = 10^{-4}$\myr) is sustained for a longer
period of time, the release of gravitational energy maintains the envelope
expansion.  After $\sim 240$\,yr, the Kelvin Helmholtz timescale has
decreased to $\sim 60$\,yr and accretion energy can now be efficiently
radiated away. At that time, the nuclear energy production is at a minimum
and the HBS is almost extinguished (Fig. \ref{MY404_07}).
The luminosity profile is now monotonicly increasing and contraction
resumes. When the accretion process ends (for an accreted mass
$M_{\mathrm{acc}} = 0.0594\Msun$ in this simulation), the contraction rate
accelerates as the envelope is no longer supported by the deposition of
accretion energy. The temperature increases and $\mathrm{L_{He}}$ rises
first, followed by $\mathrm{L_H}$. When the nuclear sources provide $\ga
90$\% of the stellar luminosity the contraction stops. The radius is then
minimal and its value is smaller than in case A, because in this
simulation, the stellar mass and thus the gravitational pull are
larger. Finally, as the temperature increases the thermal instability sets
in, but somewhat later than in case A.

\begin{figure}
\psfig{file=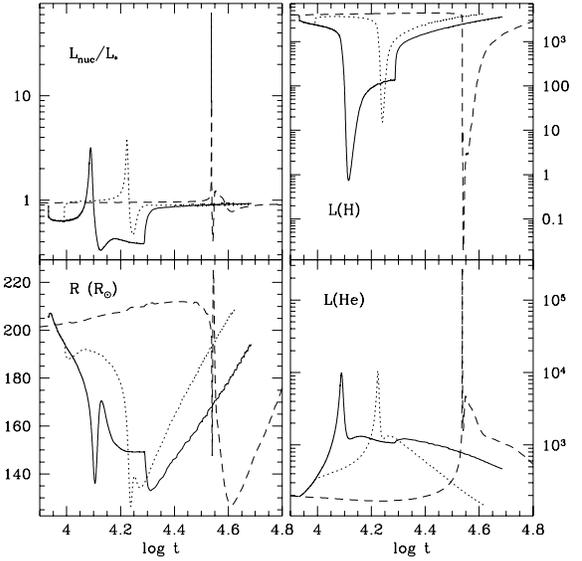,width=\columnwidth}
\caption[]{Same as Fig. \ref{MY404_07} but for low accretion rates
($10^{-5}$\myr).  The solid curves correspond to $M_{\mathrm{acc}} =
0.1082$\msun, the dotted lines to case B ($M_{\mathrm{acc}} =
0.01238$\msun) and the dashed lines represent the standard model.}
\label{MY404_08}
\end{figure}

For lower accretion rates ($\Macc = 10^{-5}$\myr), the persistent addition
of mass, up to $M_{\mathrm{acc}} = 0.1082\Msun$, increases the contraction
rate. The star heats up faster and the thermal instability develops even
earlier than in case B (Fig.  \ref{MY404_08}).
For the reasons explained previously (see Section \ref{Sect1}), the value
of $\mathrm{L_{He}}$ during the pulse is lower than in the standard
case. We also note that accretion does not perturb the thermal pulse, which
indicates that in this case this process represents a small perturbation to
the stellar structure. When the instability is over, the star resumes
contraction. The nuclear luminosity ($L_{\mathrm{nuc}}$) is lower than in
the standard model because a fraction of the energy is provided by the
accretion process.  Finally, when accretion ends, the star relaxes and
returns to a standard evolution (with a larger mass however).

To summarize these simulations, for high accretion rates the accretion
luminosity dominates the stellar energetics, reducing the nuclear energy
production of the HBS. As a result, the (nuclear) evolution of the star is
slowed down. For lower accretion rates (but still high compared to the
normal core growth rate), accretion shifts the internal structure to that
corresponding to higher masses at a faster rate than in a standard
evolution, and consequently the evolution is accelerated.
However, these results do not imply that for some intermediate value
of \macc the evolution of the star remains unchanged. Indeed, the stellar
mass increases and the core's growth rate remains always higher than in a
standard evolution. In the limiting case where the release of accretion
energy exactly counterbalances the gravitational attraction,
instantaneously the evolution of the star would remain the same. However this
situation is unstable since $M_{\mathrm{core}}$ increases. Quite rapidly the
gravitational pull takes over and the contraction resumes. Note that
for large accretion rates when the accretion process is maintained for a
longer period of time, contraction finally resumes once thermal equilibrium is
reached.

\section{Surface abundances and nucleosynthesis}
\label{nucleo}

Chemical modifications resulting from mass accretion are expected if either
a substantial amount of accreted matter is deposited in the envelope or/and
if nuclear burning is operating in the convective envelope. To separate
these two effects, we define new variables $R_1$ and $R_2$ for each
chemical element $i$ by
\begin{eqnarray}
R_{1,i} & = & \frac{\tilde X_i(t)}{X^s_i(t)} \\
R_{2,i} & = & \frac{\tilde X_i(t)}{\displaystyle
\frac{X^s_i(t)+m_{\mathrm{acc}}\,X_i^a}{1+m_{\mathrm{acc}}}}\ ,
\end{eqnarray}
where $\tilde X_i(t)$ is the mass fraction of element $i$ in an
evolutionary scheme with accretion and $X^s_i(t)$ is the mass fraction of
this element in standard computations. $m_{\mathrm{acc}}$ represents the
ratio of the total mass deposited in the envelope during the accretion
process ($ M_{\mathrm{acc,env}}$) to the mass of this region
($M_{\mathrm{env}}$). The variable $R_1$ describes the effect of accretion
on the surface abundances, including mass deposition and nuclear
burning. The variable $R_2$ represents the nuclear activity: it gives the
ratio of the mass fraction computed in our models with accretion to the
mass fraction expected from calculations with accretion if no nuclear
burning took place. Therefore, if the element is nuclearly produced then
$R_2>1$ and if it is destroyed $R_2<1$.

\begin{figure}
\psfig{file=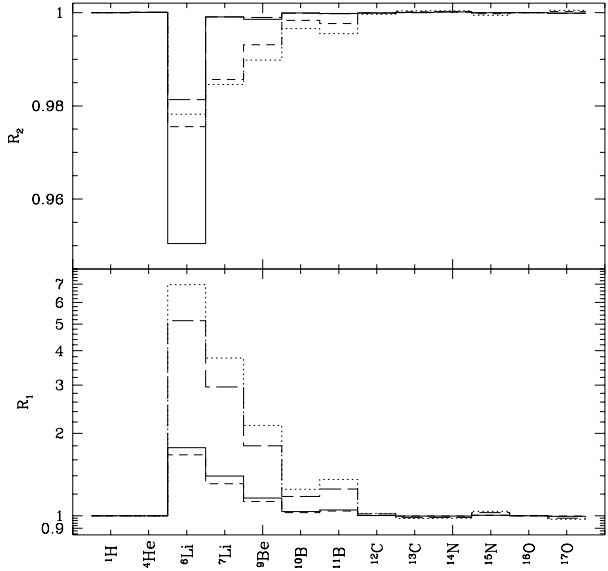,width=\columnwidth}
\caption[]{The effects of accretion on the surface abundances of several
chemical elements. The lower panel shows the ratio of the mass fraction of
an element resulting from our computations with accretion to the standard
value. The solid and short-dashed lines correspond to cases A and B,
respectively. The dotted and long-dashed lines correspond to longer
accretion phase, with accretion rates (accreted masses) that are equal to
$10^{-4}$ (0.0594\msun) and $10^{-5}$\myr (0.1082\msun), respectively. The
upper panel represents the nuclear activity. If $R_2>1$ ($R_2<1$) the
element is nuclearly produced (destroyed, see text).}
\label{MY404_09}
\end{figure}

The deposition of fresh, nuclearly unaltered material in the convective
envelope entails modifications of the surface chemical abundances.  This
effect accounts for the large values of $R_1$ attached to light elements
such as \chem{6}Li, \chem{7}Li, \chem{9}Be, \chem{10}B or \chem{11}B
(Fig. \ref{MY404_09}). 
Indeed, due to their low burning temperature, these
elements were previously burned during the pre-main sequence phase or
during the first dredge up. Depending on the mass added to the envelope,
the abundances of these elements can be increased by a factor of $\sim 2-8$
($0.3-0.9$ dex). The more mass is accreted, the higher the ratio $R_1$
is. However, the detection of these chemical changes may be difficult if
the accreted mass is small. For example, the \chem{7}Li mass abundance in
cases A and B ($M_{\mathrm{acc,env}} \sim 0.01$\msun) reaches values of
log\,$\epsilon$(Li)=1.54 and log\,$\epsilon$(Li)=1.51 respectively, which
are very close to the log\,$\epsilon$(Li)=1.40 standard value. It is
interesting to note that the accretion of planetary material in the
envelope of red giant stars was originally suggested by Alexander (1967) to
explain the high Li abundance observed in some red giants. Brown et al.
(1989) showed that a sufficient mass of terrestrial-like planets (which
have lost their initial H and He content) may replenish the envelope of
giants sufficiently to the reproduce high observed Li abundances. Note,
finally, that the surface abundances of heavier elements, with $A>13$, are
close to their initial values and are thus not modified by the accretion
process.

The upper panel of Fig. \ref{MY404_09} indicates which elements undergo
efficient nuclear reactions. The light elements are the principal target of
nuclear burning. Indeed the ``low'' temperature at the base of the
convective region, between $3\,10^6$ and $10^7$K (depending on the presence
or the absence of HBB), allows the destruction of these species.  The
nuclear burning acts preferentially on \chem{6}Li, \chem{7}Li and
\chem{9}Be. We can also detect the effects of CNO reactions in the envelope
through the production of \chem{13}C, \chem{14}N and \chem{17}O and the
depletion of \chem{12}C. However, the influence of H burning through the
CNO cycle is very small and it only appears if accretion is maintained for
a sufficiently long period of time (dotted line, dashed line in
Fig. \ref{MY404_09}).

To summarize, the temperature at the base of the convective envelope ($\la
10^7$K) is too low to give rise to a rich nucleosynthesis. The main
chemical changes thus result from the deposition of fresh material in
the convective envelope.

\section{Discussion}
\label{discuss}

Our computations involved a relatively young AGB star (at the beginning of
the thermally pulsing phase) and one may wonder what would be the effects
of brown dwarf accretion inside more evolved stars. In this respect we
note, that as the evolution progresses, the core grows but the temperature
at the base of the convective zone remains almost constant, close to
$4\,10^6$K. Thus, the location of the accretion process may remain
unchanged. However, in older AGB stars the stellar luminosity is
substantially higher and this will most certainly result in a different
stellar response to high accretion rates. Indeed, as we have seen in the
present work, the main reason for the stellar expansion is the release of a
large amount of accretion energy ($\Lacc > L_*$). We thus expect that if
mass is injected in more luminous AGB stars, then the stellar radius and
luminosity will decrease during the process, as in case B (see also Harpaz
\& Soker 1994).

Our calculations revealed that for high mass injection rates ($\Macc =
10^{-4}$\myr) the radius and luminosity of the star increase by a factor of
$\sim 2$. This leads to a substantial increase in the mass loss rate, since
according to the Reimers formula (1975) $\dot M_{\mathrm{loss}} \propto LR/M$. The
ejected mass can be estimated as 
\begin{eqnarray}
M_{\mathrm{eject}} & \sim & 10^{-4} \,\Bigl( \frac{L}{5000\Lsun}\Bigr)
\Bigl(\frac{R}{200\Rsun}\Bigr)  \Bigl( \frac{M}{\Msun}\Bigr)^{-1} \nonumber
\\  & \times & 
\Bigl(\frac{t_{\mathrm{acc}}}{100 \mbox{yr}}\Bigr)\  \Msun\ ,
\label{shell}
\end{eqnarray}
where $t_{\mathrm{acc}}$ is the duration of the accretion process. We can therefore
expect the formation of shells, corresponding to individual accretion
events. The mass of these shells is of the order of $\sim 10^{-3} -
10^{-4}$\msun.

A phenomenon not directly addressed by our spherically symmetric
calculations is that of the deposition of angular momentum into the giant's
envelope, by the accreted brown dwarf. We can however estimate the effects
of such a deposition as follows. If we assume that the secondary deposits
all of its orbital angular momentum in the envelope, and we further assume
that the envelope rotates rigidly (due to the fact that it is convective,
and thus viscous) then the ratio of the envelope angular velocity $\omega$
to the critical velocity (surface keplerian angular velocity) $\omega_K$ is
given by
\begin{equation}
\frac{\omega}{\omega_K} \approx 0.085 \,
\Bigl(\frac{M_{\mathrm{bd}}}{0.01\,M_{\mathrm{env}}} 
\Bigr) \Bigl(\frac{a}{R}\Bigr)^{1/2}\ ,
\label{omega}
\end{equation}
where $a$ is the initial orbital radius of the brown dwarf. Here we
calculated the moment of inertia of the envelope for our model, and it is
given by $I_{\mathrm{env}}=k^2 M_{\mathrm{env}}R^2$, where $k^2 \simeq
0.12$. As we can see, Eq. \ref{omega} indicates that for typical brown
dwarf masses, the AGB star can be spun-up to a significant fraction of the
critical velocity. Notably, the accretion of angular momentum is required
to explain the abnormally high rotational velocities of FK Comae stars
(e.g. Rucinski 1990, Welty \& Ramsey 1994) and of peculiar AGB stars such
as V Hydrae (Barnbaum et al. 1995).  Finally, this can also have important
consequences for the shaping of planetary nebulae (see e.g. Soker
1996a). In particular, it has been suggested that the spin-up of the
envelope is responsible for the high fraction of non-spherical planetary
nebulae (see Livio 1997 for a review).

Concerning peculiar abundances, the discovery of lithium rich G and K giant
stars (e.g. Wallerstein \& Sneden 1982, Brown et al. 1989, de la Reza \& da
Silva 1995) posed the problem of the origin of this element in these
stars. Indeed, standard stellar evolution predicts that Li should be
depleted during the pre-main sequence phase or during the first dredge
up. Several mechanisms have been proposed to explain the high Li abundance,
these include : (1) Li production (Cameron \& Fowler 1971), (2) the
retainment of the original lithium, (3) the external enrichment by novae
(e.g. Gratton \& D'Antona 1989), (4) the engulfing of a planet (e.g. Brown
et al. 1989). Recently, de la Reza et al. (1996, 1997) pointed out that
almost all the Li K giants are optical counterparts of IRAS sources, the
far IR excesses being attributed to the presence of a dusty circumstellar
shell (CS).  Based on this apparent correlation between the IR and Li
excesses, these authors proposed a scenario in which all the K giants
with masses between 1.0 and 2.5 \msun\, become Li rich after the first
dredge-up. They assume that the mechanism responsible for the Li enrichment
is accompanied by a sudden mass loss which leads to the formation of a
CS. The Li which has not been ejected remains in the stellar surface and is
depleted later when the mass loss stops. In this scenario, the CS expands,
cools down, disappears and the Li surface abundance reaches again standard
values.  This evolutionary path, or ``Li path'', can reasonably explain the
Li abundance distribution of K giant stars in a colour-colour diagram. Its
duration is of the order of $\sim 10^5$\,yr and it requires that Li suddenly
appear at the surface of the star and then be depleted on a timescale of $
10^3 - 10^5$\,yr. However, de la Reza et al. do not provide convincing
mechanisms for Li enrichment and mass loss. They invoke an extra mixing
process for Li production, the ``cool bottom processing'' (Wasserburg,
Boothroyd \& Sackmann 1995), but they remain vague concerning the
generation of mass loss. We propose here that the accretion of a planet or
a brown dwarf by a giant star could effectively explain all of these
features. Indeed, our calculations show that the accretion process (1) can
increase the mass loss rate and lead to the ejection of a shell
(Eq. \ref{shell}), and (2) can enrich the envelope with lithium
(Fig. \ref{MY404_09}), in agreement with observations.

Observations also show that lithium rich giants have generally normal
rotational velocities [de Medeiros et al. 1996; although some rapid
rotators exist, e.g. 1E 1751+7046 (Ambruster et al. 1997)]. This implies
that if the accretion of a planet/brown dwarf is responsible for the Li
enhancement, some fraction of the orbital angular momentum which is
deposited into the envelope has to be removed (or that in most cases of
engulfed planets $a/R<1$, see Eq. \ref{omega}). This can be achieved by
efficient magnetic braking, since the giants develop a deep convective
envelope (e.g. Leonard \& Livio 1995). We should note that the paucity of
giants showing both rapid rotation and high Li abundance probably indicates
that the timescale for rotational braking is short compared to the
giant evolutionary phase (a few $10^7$\,yr).

The existence of very rich Li stars (e.g. Gregorio-Hetem et al. 1992,
Torres et al. 1995, de la Reza \& da Silva 1995, Fekel et al. 1996) poses a
somewhat more difficult challenge.  One could imagine, however, that the
collision of the planet (or a brown dwarf) with the star's core may trigger
the ``cool bottom processing'' (Wasserburg, Boothroyd \& Sackmann
1995). In this scenario, \chem{7}Be-rich material can be dredged-up from
the deep interior to the convective envelope where it decays through the
\chem{7}Be($e^-,\nu$)\chem{7}Li reaction.

A comment should be added about statistics. From a sample of 644 stars,
Brown et al. (1989) found that $\sim 4$\% of G and K giants have Li
abundances above $\log \epsilon({\mathrm Li})\sim 1.3$ and that an
additional 4\% have abundances between 1.2 and 1.3. Considering the fact
that many of the Li rich stars show an IR excess compatible with the
presence of a surrounding shell, it is tempting to interpret these
observations as a signature of an accretion event. This could indicate that
$4-8$\% of all stars have low-mass (possibly substellar) companions. These
statistics are fully compatible with those inferred from PNe nuclei
(e.g. Yungelson, Tutukov \& Livio 1993).

\section{Summary}
\label{summary}

We have presented computations in which a brown dwarf (or a planet) is
accreted in the core of an AGB star. This situation results when the brown
dwarf is engulfed in the stellar envelope as the latter is expanding due to
stellar evolution. Viscous friction as well as tidal effects generate a
drag force that makes the brown dwarf spiral in. We have shown that the
region where the brown dwarf is expected to disrupt is close to the base of
the convective envelope.

We investigated the response of the star to the accretion of a small body
for two accretion rates, $10^{-4}$ and $10^{-5}$\myr.  Our simulations show
that for the higher accretion rate, the main energetic effects come from the
release of potential energy. The accretion luminosity is of the order of
the stellar luminosity and this additional energy source makes the star
expand. Another interesting feature of these computations (involving a high
accretion rate) is the appearance of hot bottom burning. However, the
relatively low temperature at the base of the convective envelope ($\sim
10^7$K) and the short duration of the accretion phase do not allow for a
rich nucleosynthesis in the convective zone. We find that accretion delays
the development of the thermal instability.  The nuclear energy production
due to the H and He burning shells is smaller than in a standard case
because a part of the energy is provided by the deposition of the accretion
energy.  Hydrogen and helium are consequently burned more slowly and the
evolution is slowed down.

For the lower accretion rate the evolution is different. In this case the
accretion luminosity represents a small fraction of the stellar luminosity
and therefore expansion does not ensue. Conversely, the addition of mass
increases the gravitational pull and the star contracts even more
efficiently than in a standard scheme. The temperature of the nuclear
burning shells increases more rapidly and the thermal instability occurs
earlier than in conventional evolution. The global stellar evolution is
accelerated.

Except for variations in the luminosity and radius, both of which can be
increased by a factor of $\sim 2$, the effects of mass accretion in an
evolved star may be traced by changes in the surface chemical
abundances. We showed that such changes are mainly the result of material
deposition rather than nucleosynthesis. The chemical changes affect mainly
light elements such as \chem{7}Li and \chem{9}Be whose mass fraction can be
increased by a factor of 2-8, depending on the accretion rate and duration
of the accretion process.

Generally, the observational signatures of the accretion of a brown dwarf or
a planet by giant stars can be (depending on the physical parameters
involved): $(i)$ the formation of a circumstellar shell, $(ii)$ the
enhancement of the surface \chem{7}Li abundance, $(iii)$ rapid rotation. We
propose that the IR excesses and high Li abundances observed in $4-8$\% of
the G and K giants originate from the accretion of massive planets/brown
dwarfs/very low-mass stars. 

Finally, we should note that we also made several attempts to accrete
matter deeper inside the star (e.g. below the HBS). However, such a process
could not be followed reliably with our code. The deposition of protons in
the very central regions increases the nuclear energy production to such
rates, that the envelope cannot absorb this energy excess.

\section*{Acknowledgments}
 
LS wishes to thank Manuel Forestini for invaluable discussions and also for
kindly providing us with the initial stellar model. LS acknowledges support
from the French Ministry of Foreign Affairs (Bourse Lavoisier) and thanks
the STScI for its hospitality. This work has been supported in part by NASA
grants NAGW-2678, G005.52200 and G005.44000, and the Director's
Discretionary Research Fond at STScI.  The computations presented in
this paper were performed at the ``Centre Intensif de Calcul de
l'Observatoire de Grenoble''.

\end{document}